# Tip-Enhanced Sum Frequency Generation for Molecular Vibrational Nanospectroscopy

*Atsunori Sakurai,[1,2,3]\* Shota Takahashi,[1] Tatsuto Mochizuki,[1,2] and Toshiki Sugimoto[1,2,3]\**

[1]Institute for Molecular Science, National Institutes of Natural Sciences, Okazaki, Aichi 444-8585, Japan

[2]Graduate Institute for Advanced Studies, SOKENDAI, Okazaki, Aichi 444-8585, Japan

[3]Laser-Driven Electron-Acceleration Technology Group, RIKEN SPring-8 Center, Sayocho, Hyogo 679-5148, Japan

ABSTRACT. Vibrational sum frequency generation (SFG) is a nonlinear spectroscopic technique widely used to study the molecular structure and dynamics of surface systems. However, the spatial resolution achieved by far-field observations is constrained by the diffraction limit, obscuring molecular details in inhomogeneous structures smaller than the wavelength of light. To overcome this limitation, we developed a system for tip-enhanced SFG (TE-SFG) spectroscopy based on a scanning tunneling microscope. We successfully detected vibrational TE-SFG signals from adsorbed molecules on a gold substrate under ambient conditions. The phase analysis of interferometric SFG spectra provided information on molecular orientation. Furthermore, the observed TE-SFG signal was confirmed to originate from a highly localized region within a gap between the tip apex and the sample substrate. This method offers a novel platform for nonlinear optical nanospectroscopy, paving the way for the investigation of surface molecular systems beyond the diffraction limit.

The surfaces of materials exhibit different structural orders from the bulk, allowing various molecules and ions to interact and induce reactions. Vibrational sum frequency generation (SFG) spectroscopy is a second-order nonlinear spectroscopic technique widely used for studying the molecular structures and dynamics of surface systems because it is surface-sensitive due to symmetry requirements,[1] capable of probing molecular orientations,[2-4] and applicable to time-resolved spectroscopy.[5-7] Owing to its versatility, vibrational SFG has been employed to investigate a variety of systems, such as monolayers on a liquid surface (Langmuir films),[8, 9] self-assembled monolayers (SAMs) on metals,[10-14] and water molecular structures and dynamics at surfaces or interfaces.[15-22] Additionally, vibrational SFG has been used for in-situ observations of chemical reactions and crystal growth, including catalytic reactions,[23-25] electrochemical reactions,[26-29] and the growth of highly oriented ice.[30-32] Furthermore, the observation of microstructures using SFG in combination with microscopy is being actively pursued.[33-37] However, the spatial resolution achieved by conventional far-field observations is constrained by the diffraction limit of light. Thus, the applicability of SFG to nanospectroscopy in systems with inhomogeneous structures smaller than the wavelength of light is restricted.

To overcome this limitation, considerable progress has been made by using tip-enhanced spectroscopy (TES)[38-46] and scattering-type scanning near-field optical microscopy (s-SNOM).[47-]

[52] These techniques combine the high spatial resolution of scanning probe microscopy—including scanning tunneling microscopy (STM) or atomic force microscopy (AFM)—with the analytical capabilities of both linear and nonlinear optical spectroscopy. Recent advances in TES allow the visualization of single molecules by means of Raman scattering[53, 54] as well as photoluminescence.[55, 56] Moreover, using many different nonlinear optical processes (second-harmonic generation (SHG), SFG, two-photon photoluminescence, and four-wave mixing (FWM)), the nanoimaging of metal structures[57-59] and the elucidation of exciton dynamics in transition-metal dichalcogenides[60] have been demonstrated. The essential component for realizing such high-spatial-resolution imaging and nonlinear optical nanospectroscopy is electric field confinement and enhancement, originating from plasmon resonance in the nanogap formed between a metal tip and a metal substrate. However, plasmon resonances in nanogaps (i.e., gap-mode plasmons) typically occur in the visible and near-infrared (near-IR) regions[61-63] and do not significantly enhance the infrared (IR) signal. Thus, applications of TES are limited to the visible and near-IR regions. In contrast, IR $s$-SNOM extracts weak near-field signals using an interferometric technique[64] and lock-in detection.[65] IR $s$-SNOM has been applied to various systems, such as surface plasmons in graphene[66-68] and the insulator-metal transition in vanadium dioxide ($VO_2$).[69, 70] In particular, near-IR pump and IR probe nanospectroscopy, as well as FWM nanospectroscopy, have been used to examine the nonlinear responses of these systems, clarifying the dissipation and propagation mechanisms of graphene plasmons[71-73] and the photo-induced phase transition and carrier dynamics in $VO_2$.[74, 75] Despite their broad range of applications, neither TES nor IR $s$-SNOM has been implemented for nonlinear molecular vibrational spectroscopy, let alone vibrational SFG.

Upconverting optical responses from the mid-IR region to the visible region may leverage the significant field enhancement from gap-mode plasmons, making it advantageous for molecular nanospectroscopy since molecular vibrations exist in the mid-IR region. Taking this into account, we focus on the potential of upconversion techniques and plasmonic enhancements at the nanoscale. Recently, we studied tip-enhanced SHG (TE-SHG) spectroscopy over a wide wavelength range.[76] In this method, near-IR excitation pulses are upconverted into frequency-doubled signals in the visible region. We identified two mechanisms for SHG signal enhancement: (i) enhancement of the incident electric field, which occurs as the incident light received by the tapered part of the tip is propagated to the tip apex (antenna effect), and (ii) enhancement of the radiation efficiency from second-order polarization in the visible region, where gap-mode plasmon resonance exists (plasmon resonance). In the electromagnetic field simulations, the antenna effect was shown to increase as the wavelength of the incident light increased (Fig. 1(a)).[76] This suggests that the antenna effect works more effectively with mid-IR incident light. Meanwhile, vibrational SFG is a nonlinear optical process where the molecular vibration, resonantly excited by a mid-IR pulse, is upconverted to second-order polarization by another near-IR or visible pulse, and then sum-frequency light is radiated in the visible region. Therefore, if the sum frequency matches the plasmon resonance, it can further enhance the radiation efficiency of the SFG signal (Fig. 1(b)); however, achieving this for molecular observation remains a challenge. Thus, in the present study, we used TE-SFG spectroscopy for the detection of SFG signals from molecules. We successfully detected vibrational TE-SFG signals from the ultrathin layer of adsorbed molecules in a plasmonic nanocavity.

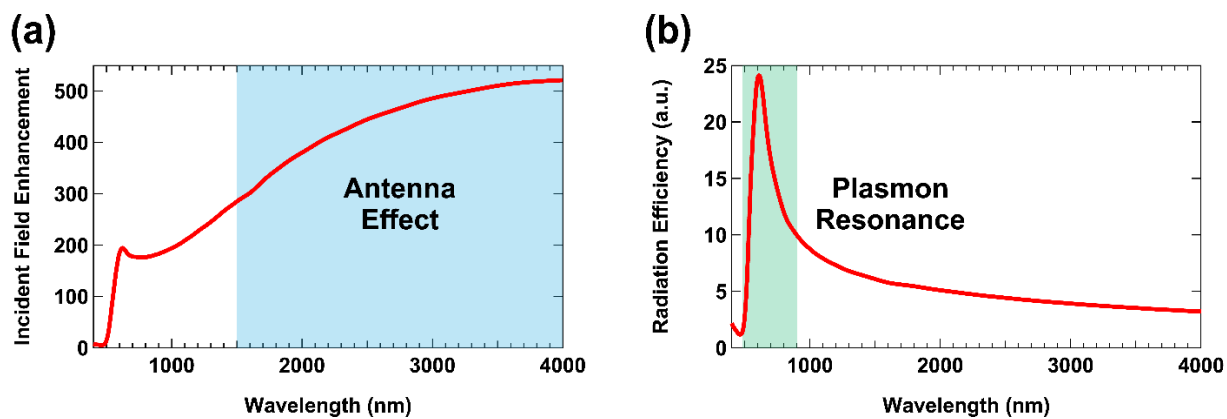

**Figure 1.** (a) Enhancement of the incident field at the nanojunction between the gold tip and the gold substrate, calculated using electromagnetic field simulations. Significant enhancement is observed as the wavelength of the incident light increases, which results from the antenna effect of the tip. Plasmon resonance also occurs in the visible region, but its enhancement is smaller than that observed for the antenna effect in the IR region. (b) Radiation efficiency from the dipole in the nanojunction, monitored at a position corresponding to the far-field. Strong enhancement of radiation efficiency is observed in the visible region due to gap-mode plasmon resonance. Details regarding the electromagnetic field simulations are described in Supporting Information §5.

Two incident beams were used: a tunable IR laser (2800–3150 cm$^{-1}$ at central wavenumber) and a narrowband 1033-nm laser (10 cm$^{-1}$ in full width at half maximum). They were collinearly focused by a lens on a nanojunction between a tip and a sample substrate mounted on an STM (Fig. 2(a)). The central wavenumber of the IR laser determined the portion of the vibrational mode that was probed. The input powers of the IR and 1033-nm lasers were 30 and 0.3 mW, respectively, with a repetition frequency of 50 MHz. The incident beams were linearly polarized along the tip axis (p-polarization) to effectively excite the gap-mode plasmon. The emitted SFG signal was collected by another lens in a direction symmetric to the surface normal (see Supporting Information §1 for details regarding the experimental setup). We used a homemade gold tip prepared from a gold wire by electrochemical etching,[77] and Fig. 2(b) shows an image obtained using scanning electron microscopy (SEM). Additional SEM images of the tip at different resolutions are presented in Supporting Information §1. The sample was an SAM of 4-methylbenzenethiol (4-MBT) on a Au(111) film evaporated on a mica substrate (see Supporting Information §1 for details regarding the sample preparation). SAM is a well-known model system for organic monolayers.[78-80] The SFG spectrum of $Au–S–C_6H_4–CH_3$ in the CH-stretch region has three prominent transitions near 2900 $cm^{-1}$ due to the terminal methyl groups.[13] A topographic image of 4-MBT on Au(111) is shown in Fig. 2(c). Observation of the Au(111) steps shows that the surface is atomically flat. Moreover, the many island-like structures (adatom islands), which are typical of SAMs of arenethiols on Au(111),[78, 80] confirm the formation of a single monolayer of 4-MBT. In our STM setup, the tip was fixed such that incident beams irradiated the tip apex during sample scanning. When incident beams were irradiated, the tip‐sample distance was kept constant by setting the tunneling current at 2 nA with a bias voltage of 0.5 V. To avoid sample damage during the measurement, the sample was scanned over a 3 μm × 3 μm area at a scanning speed of 2.5 μm/s. All experiments were performed under ambient conditions.

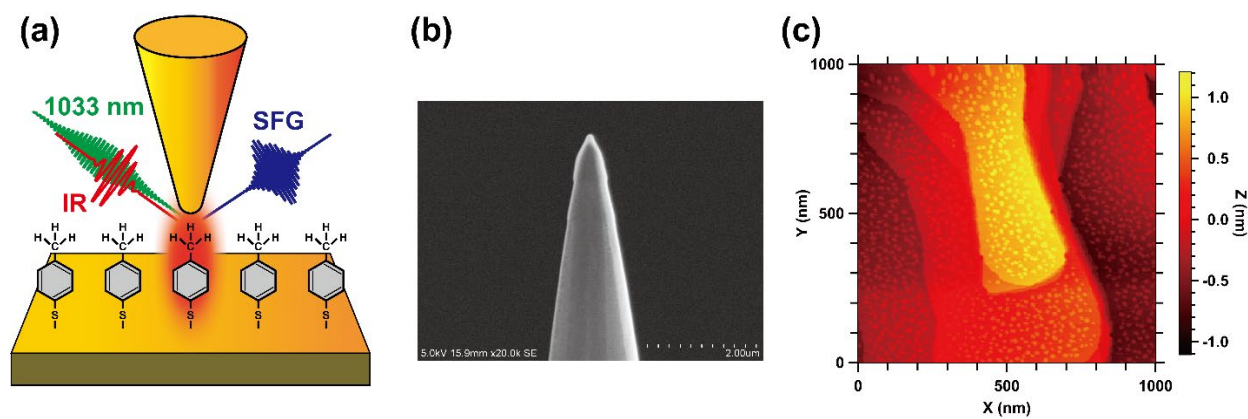

**Figure 2.** (a) Schematic of TE-SFG spectroscopy. Tunable IR and narrowband 1033-nm laser beams were collinearly focused on the nanojunction between the STM tip and sample substrate. (b) SEM image of the gold STM tip used in the experiment. (c) STM topographic image of an SAM of 4-MBT on Au(111). The scanned area was 1000 nm × 1000 nm (256 pixels × 256 lines), and the color scale bar indicates the height. The tunneling current was set at 0.5 nA with a bias voltage of 0.1 V.

Figs. 3(a) and (b) show the obtained time series of SFG spectra with IR central wavenumbers at 3050 and 2900 cm$^{-1}$, respectively. Despite the movement of the tip contact position on the sample surface during measurement, the time variation in the SFG spectra was small. This appeared to be due to the homogeneous nature of the sample surface and the averaging of environmental thermal fluctuations over the exposure time. By integrating the time series of SFG spectra, the turquoise blue curve in Fig. 3(c) and the dark yellow curve in Fig. 3(d) were obtained, which are labeled "Tunneling." We also measured spectra with the sample substrate retracted by 50 nm (represented by the black curves in Figs. 3(c) and (d), labeled "Non-contact").

The SFG signal intensity significantly increased when the sample was sufficiently close to the tip, where the tunnel current was detectable. This increase in the SFG signal arose from the near-field enhancement caused by the gap-mode plasmon, i.e., TE-SFG. In Fig. 3(c), in the region where 4-MBT does not have vibrational resonance, the TE-SFG spectrum has a shape similar to that of the IR pulse spectrum but is slightly broadened. This SFG signal is attributed to the vibrationally non-resonant responses of gold,[10, 27] and the spectrum represents the convolution of the IR pulse with a narrowband 1033-nm pulse.[81] Consequently, the SFG spectrum becomes slightly broader than the IR pulse spectrum. In contrast, the TE-SFG spectrum in Fig. 3(d) exhibits a characteristic dip structure. Furthermore, similar dip structures are observed at 2850 and 2940 cm$^{-1}$ (Fig. 3(e)). Such dip structures in SFG spectra have been widely reported for SAM molecules on gold surfaces in far-field measurements.[10, 11, 13, 14, 82] These spectral features result from the interference between the vibrational resonant responses of molecules and the vibrational non-resonant response of gold.[1, 10-14, 82] The spectra provide evidence for observing vibrational TE-SFG signals from 4-MBT molecules in the nanogap. As described in Supporting Information §2, these spectral shapes are approximated by the following equation:[12, 20, 25, 27, 83-85]

$$I_{\text{SFG}}(\omega_{\text{IR}}) = c \exp\left[-\frac{(\omega_{\text{IR}} - \omega_0)^2}{\sigma^2}\right] \left|1 + \sum_q \frac{r_q e^{-i\theta}}{\omega_q - \omega_{\text{IR}} - i\Gamma_q}\right|^2, \tag{1}$$

where $\omega_{\text{IR}}$ is the frequency in the IR region; $c$ is the coefficient of relative strength; $\omega_0$ and $\sigma$ represent the central frequency and spectral width of the incident IR pulse, respectively; $\theta$ is the phase factor of the vibrational non-resonant contribution; and $r_q$, $\omega_q$, and $\Gamma_q$ represent the amplitude relative to the vibrational non-resonant contribution, the resonant frequency, and the damping constant of the $q$-th vibrational mode, respectively.

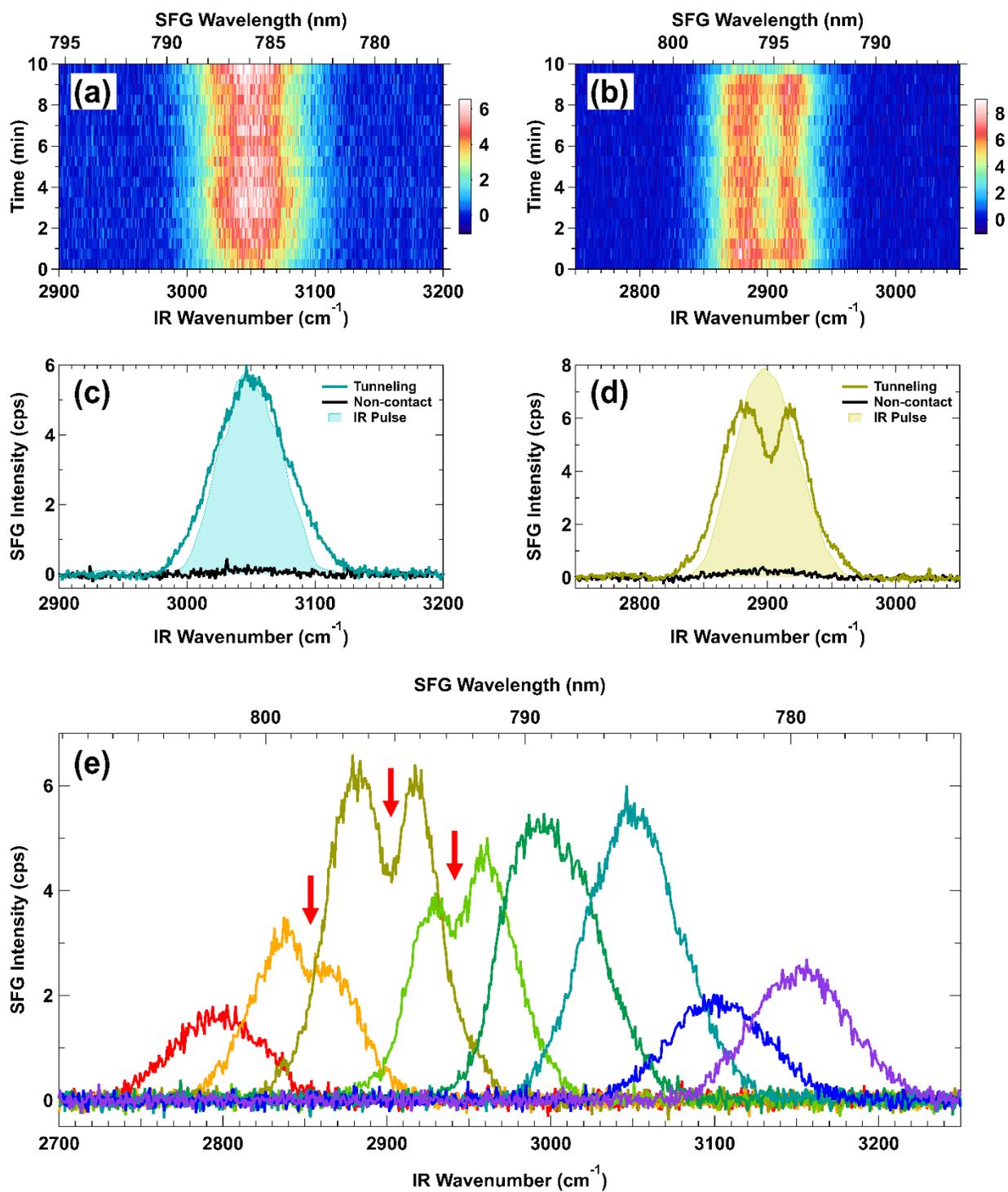

**Figure 3.** Time series of SFG spectra at IR central wavenumbers of (a) 3050 cm⁻¹ and (b) 2900 cm⁻¹. The exposure time for each spectrum was 30 s, and the measurement was repeated 20 times. The bottom axis corresponds to the IR wavenumber, and the top axis corresponds to the SFG wavelength. The intensity is normalized by the exposure time as counts per second (cps). The turquoise blue curve in (c) and the dark yellow curve in (d) represent the SFG spectra obtained from (a) and (b), respectively, by integrating them over time (Tunneling). The black curves represent the spectra obtained with the sample retracted by 50 nm (Non-contact). For reference, the incident IR pulse spectra are superimposed. (e) TE-SFG spectra for different IR central wavenumbers ranging from 2800 to 3150 cm$^{-1}$ at intervals of 50 cm$^{-1}$. The spectra obtained with the sample substrate retracted by 50 nm were subtracted as the background in each case. Dip structures are observed at the positions indicated by the red arrows.

To extract information on the molecular vibrations, we smoothed the TE-SFG spectra and performed fitting using Eq. (1), assuming a single vibrational mode within each spectral range. The fitting results are indicated by the black dashed curves in Fig. 4, which agree well with the experimental results. The red curves represent the imaginary part of the resonant contribution in the second-order susceptibility ($\mathrm{Im}\big[\chi_{\mathrm{R}}^{(2)}(\omega_{\mathrm{IR}})\big]$), calculated using the same parameters as the fitting results. These curves correspond to the purely absorptive response of the $q$-th vibrational mode.[4, 21]

$$\mathrm{Im}\big[\chi_{\mathrm{R}}^{(2)}(\omega_{\mathrm{IR}})\big] = \mathrm{Im}\left[\frac{r_q}{\omega_q - \omega_{\mathrm{IR}} - i\Gamma_q}\right] = \frac{r_q \Gamma_q}{\left(\omega_q - \omega_{\mathrm{IR}}\right)^2 + \Gamma_q^2}. \qquad (2)$$

In Figs. 4(a–c), the best-fitting frequencies of the vibrational modes are 2853, 2904, and 2939 cm$^{-1}$, which correspond to a methyl symmetric stretching (CH$_3$-SS) vibration, a Fermi resonance (CH$_3$-FR) between CH$_3$-SS and an overtone of a methyl bending vibration, and a degenerate methyl stretching (CH$_3$-DS) vibration, respectively.[10, 13] These values agree well with the frequencies observed in the far-field SFG measurements for the same sample (Supporting Information §3). The negative signs of $\text{Im}\left[\chi_R^{(2)}(\omega_{IR})\right]$, resulting from the negative signs of $r_q$ (Table S2, Supporting Information §6), indicate that the terminal methyl groups are oriented away the gold surface,[4, 82, 86] which is consistent with the adsorbed structures of the 4-MBT molecules. Moreover, the phase of the vibrationally non-resonant SFG signal from gold ($\theta$) is fitted to approximately 90° (Table S2), matching previously reported values,[12, 82] as well as our far-field measurement result (Table S1). The best-fitting parameters are presented in Table S2. Despite the influence of plasmonic field enhancement, the TE-SFG spectra can be interpreted similarly to far-field observations, as explained in Supporting Information §4.

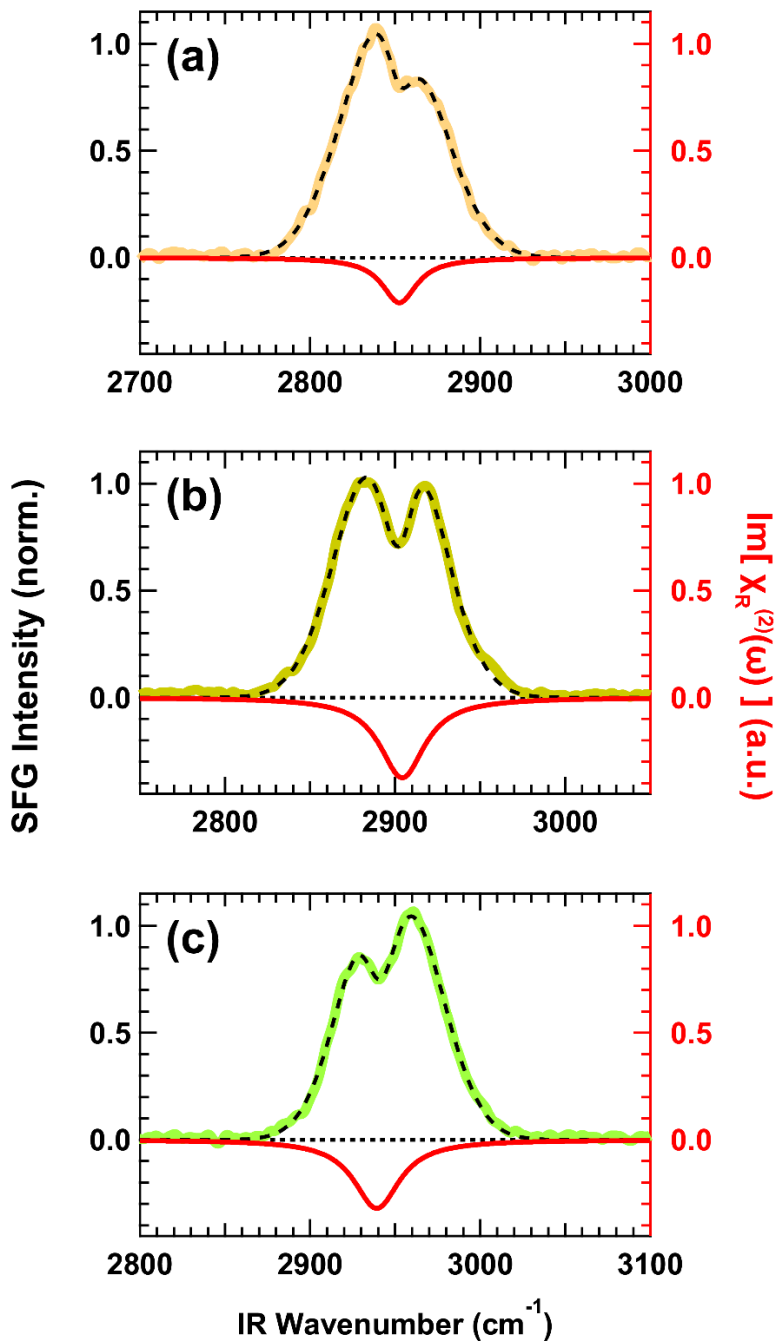

**Figure 4.** TE-SFG spectra in Fig. 3(e) were smoothed and fitted using Eq. (1). The fitting results are represented by the black dashed curves. The red curves represent the imaginary part of the resonant contribution in the second-order susceptibility ($\mathrm{Im}\big[\chi_R^{(2)}(\omega_{IR})\big]$), which corresponds to the purely absorptive vibrational responses given by Eq. (2). The extracted peaks correspond to

(a) $CH_3$-SS at 2853 $cm^{-1}$, (b) $CH_3$-FR at 2904 $cm^{-1}$, and (c) $CH_3$-DS at 2939 $cm^{-1}$. The left axis corresponds to the normalized intensity of SFG, and the right axis corresponds to the intensity of $Im\left[\chi_R^{(2)}(\omega_{IR})\right]$, with both axes having the same scale.

Next, we investigated the localization of the signal-generation region in TE-SFG near the tip apex. The TE-SFG signal was observed only when the tip was close to the sample and disappeared when the sample was retracted by 50 nm (Figs. 3(c) and (d)); this suggests that the signal-generation region of TE-SFG is confined to a region smaller than 50 nm. To evaluate the extent of optical field confinement, we measured the TE-SFG spectra at different tip–sample distances (Fig. 5(a)) and plotted the integrated counts of the vibrational non-resonant SFG signals within the range of 2820–2980 $cm^{-1}$ as a function of the relative tip–sample distance (Fig. 5(b)) (see Supporting Information §8 for procedural details). The results indicate a significant difference in TE-SFG intensity between the original position and the position after the sample was retracted by approximately 1 nm. Note that the tip–sample distance is determined by the tunneling current. Laser irradiation induces electron tunneling driven by plasmonic field enhancement (plasmon-assisted tunneling), which increases the tunneling current.[63, 87] Therefore, even if the tunneling current is set to a specific value, the actual tip–sample distance under laser irradiation may be longer than expected. This effect becomes pronounced when the tip–sample distance decreases. Plasmon-assisted tunneling induced fluctuations in the tip–sample distance, which were measured to be 0.42 and 0.54 nm for the relative tip–sample distances of 0 and 0.98 nm, respectively, as detailed in Supporting Information §9. The error bars on the horizontal axis at these two points are based on the measured fluctuations (Fig. 5(b)). For the other relative tip–sample distances, the error bars present the average of these two values for visual reference. In addition, the error bars on the vertical axis are derived from the standard deviation of 12 repeated spectra at each relative tip–sample distance (see Supporting Information §10). From these results, even taking the effect of plasmon-assisted tunneling into account, the signal-generation region is considered to be localized within a gap of approximately 1 nm between the tip apex and the substrate. This localization is facilitated by the optical field confinement through the gap-mode plasmon, generated by the up-conversion process from the IR region to the visible region.

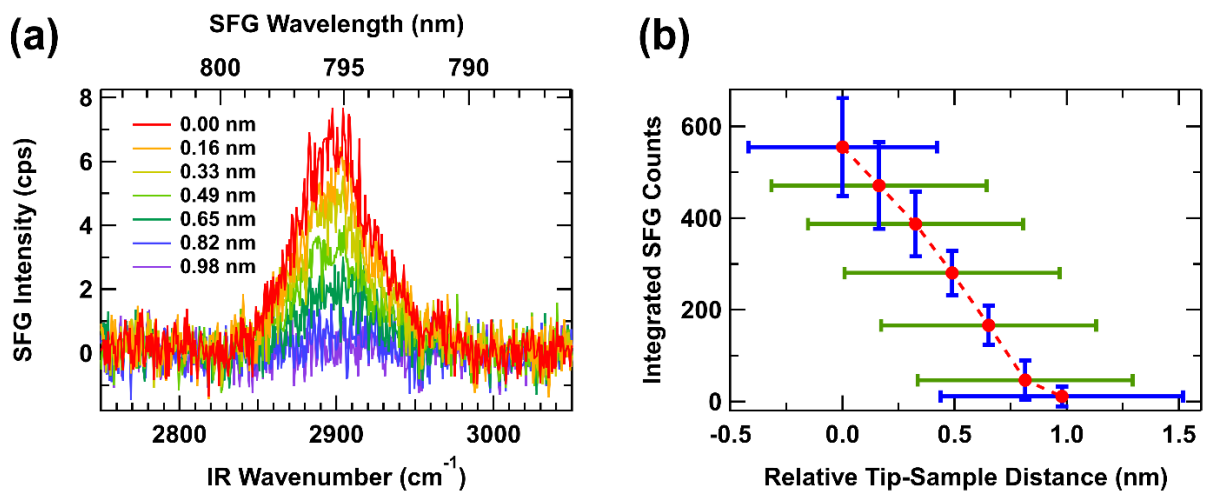

**Figure 5.** (a) TE-SFG spectra for different relative tip–sample distances, which was varied from 0.00 to 0.98 nm in increments of 0.16 nm. The spectrum obtained at the relative tip–sample distance of 50 nm was subtracted as the background in each case. (b) Approach curve of TE-SFG. Integrated SFG counts within the range of 2820–2980 cm$^{-1}$ are plotted as a function of the relative tip–sample distance. Details regarding the determination of the error bars are presented in the main text.

With regard to the localization of signal-generation regions, we also consider the differences in signal enhancement between TES and surface-enhanced spectroscopy (SES). Both TES and SES rely on localized surface plasmons: SES utilizes plasmonic structures such as rough metal surfaces or metal nanoparticles, which enhance the electric field at multiple hot spots. In contrast, TES exploits the gap-mode plasmon formed between the metallic tip apex and substrate, forming a single hot spot. Both methods have been successfully applied to Raman scattering as tip-enhanced Raman spectroscopy and surface-enhanced Raman spectroscopy (SERS), achieving significant signal enhancements of up to $10^6$–$10^{14}$.[88, 89] However, applying SES to SFG is not straightforward. In fact, the signal enhancement in SFG for the same material (4-MBT) generated by the gaps between gold nanoparticles and a gold substrate was reported to be significantly smaller, with a maximum enhancement factor of 5.[14] This was attributed to the coherent nature of the SFG process, where the electric fields from molecular polarizations in each hot spot interfere destructively when the hot spots are randomly positioned,[14] in contrast to SERS, where the incoherent Raman scattering signals are simply summed. In our measurements, the signal-generation region was confined to a single small nanogap region, which prevented destructive interference. Consequently, strong signal enhancement in SFG, along with the extraction of phase information, was achieved in our TE-SFG measurements.

In summary, we successfully observed vibrational SFG signals from adsorbed molecules beyond the diffraction limit using tip enhancement. The three dip structures observed in the SFG spectra were assigned to the CH-stretch modes of the methyl group in 4-MBT. The phase analysis of the second-order susceptibility provided information on molecular orientations consistent with the adsorbed structures of the 4-MBT molecules. The approach curve of the TE-SFG indicated that the signal-generation region was highly localized within a gap of approximately 1 nm between the tip apex and the substrate. This suggests that TE-SFG has the potential to realize high spatial resolution and high signal sensitivity for single-molecule detection. In addition, all measurements were conducted under ambient conditions, rather than in an ultra-high-vacuum and low-temperature environment, which indicates that the TE-SFG technique is applicable not only to model systems but also to practical surface systems, such as noble metal catalysts[90, 91] or electrode surfaces.[92]

ASSOCIATED CONTENT

**Supporting Information**.

The following files are available free of charge.

 Methods, including the experimental setup, spectra of incident lasers, sample preparation, and SEM images of the Au tips; interpretation of vibrational SFG spectra; far-field SFG measurements and analysis; enhancement mechanism in TE-SFG spectroscopy; numerical procedure for electromagnetic field simulations; best-fitting parameters used in Figure 4; confirmation of the reproducibility of dip structures in TE-SFG spectra; measurement procedure for the approach curve of TE-SFG; temporal variation in tip–sample distance and tunneling current before and during laser irradiation; intensity fluctuations in TE-SFG signals for each relative tip–sample distance  (PDF)

AUTHOR INFORMATION

**Corresponding Authors**

*__Atsunori Sakurai__ — Institute for Molecular Science, National Institutes of Natural Sciences, Okazaki, Aichi 444-8585, Japan; Graduate Institute for Advanced Studies, SOKENDAI, Okazaki, Aichi 444-8585, Japan; Laser-Driven Electron-Acceleration Technology Group, RIKEN SPring-8 Center, Sayocho, Hyogo 679-5148, Japan

https://orcid.org/0000-0002-4410-7518

Email: asakurai@ims.ac.jp

*__Toshiki Sugimoto__ — Institute for Molecular Science, National Institutes of Natural Sciences, Okazaki, Aichi 444-8585, Japan; Graduate Institute for Advanced Studies, SOKENDAI, Okazaki, Aichi 444-8585, Japan; Laser-Driven Electron-Acceleration Technology Group, RIKEN SPring-8 Center, Sayocho, Hyogo 679-5148, Japan

https://orcid.org/0000-0003-3453-6009

Email: toshiki-sugimoto@ims.ac.jp

**Authors**

__Shota Takahashi__ — Institute for Molecular Science, National Institutes of Natural Sciences, Okazaki, Aichi 444-8585, Japan

https://orcid.org/0000-0002-7191-5051

__Tatsuto Mochizuki__ — Institute for Molecular Science, National Institutes of Natural Sciences, Okazaki, Aichi 444-8585, Japan; Graduate Institute for Advanced Studies, SOKENDAI, Okazaki, Aichi 444-8585, Japan

https://orcid.org/0000-0002-8587-3585

**Author Contributions**

**Funding Sources**

T.S. acknowledges financial support from JST-PRESTO (JPMJPR1907); JST-CREST (JPMJCR22L2); JSPS KAKENHI Grants-in-Aid for Scientific Research (A) (19H00865 and 22H00296); and the ATLA Innovative Science and Technology Initiative for Security (JPJ004596). A.S. acknowledges financial support from JSPS KAKENHI Grants-in-Aid for Scientific Research (B) (23H01855); Grants-in-Aid for Early-Career Scientists (20K15236); the Casio Science Promotion Foundation (38-06); and the Research Foundation for Opto-Science and Technology. S.T. acknowledges financial support from Grants-in-Aid for JSPS Fellows (22KJ3099).

**Notes**

The authors declare no competing financial interest.

ACKNOWLEDGMENT

A.S. is grateful to Jun Nishida, Takashi Kumagai, and Akihito Ishizaki at the Institute for Molecular Science (IMS) for the insightful discussions, helpful advice, and constructive comments on this manuscript. We thank the members of the Equipment Development Center of IMS: Masaki Aoyama, Takuhiko Kondo, Nobuo Mizutani, and Takuro Kikuchi for their technical assistance with machine work; as well as Tomonori Toyoda and Kazunori Kimura for their technical assistance with electronics. We also thank Aya Toyama and Osamu Ishiyama at the Instrument Center of IMS for their tutorial on SEM. SEM observation of tips was conducted at IMS, supported by "Advanced Research Infrastructure for Materials and Nanotechnology in Japan (ARIM)" of the Ministry of Education, Culture, Sports, Science and Technology (MEXT), Proposal Number JPMXP1223MS5022.

# Supporting Information: Tip-Enhanced Sum Frequency Generation for Molecular Vibrational Nanospectroscopy

*Atsunori Sakurai,[1,2,3]\* Shota Takahashi,[1] Tatsuto Mochizuki,[1,2] and Toshiki Sugimoto[1,2,3]\**

[1]Institute for Molecular Science, National Institutes of Natural Sciences, Okazaki, Aichi 444-8585, Japan

[2]Graduate Institute for Advanced Studies, SOKENDAI, Okazaki, Aichi 444-8585, Japan

[3]Laser-Driven Electron-Acceleration Technology Group, RIKEN SPring-8 Center, Sayocho, Hyogo 679-5148, Japan

\*Corresponding authors: asakurai@ims.ac.jp, toshiki-sugimoto@ims.ac.jp

Table of Contents

# 1. Methods

## Experimental setup.

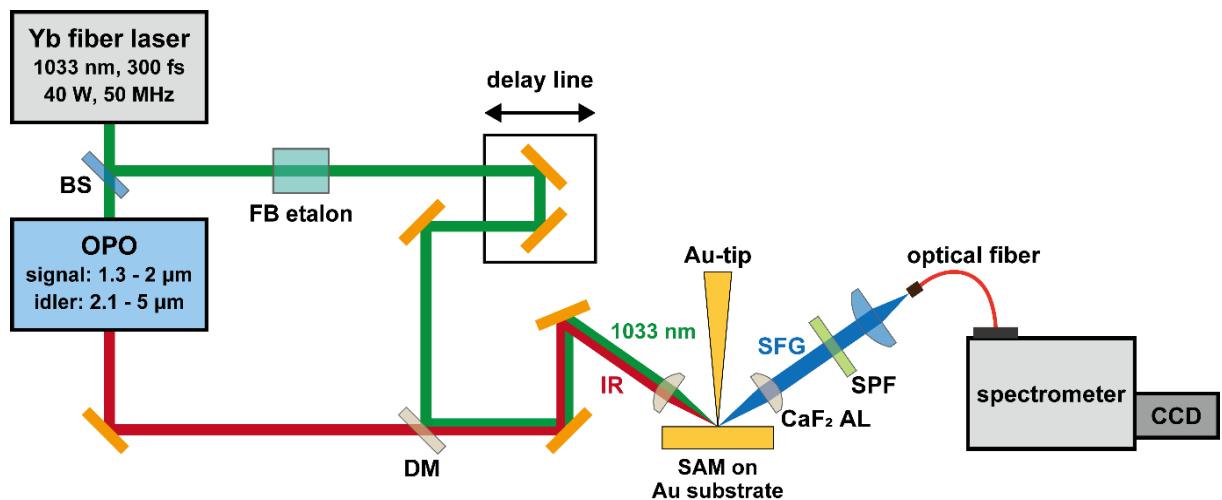

**Figure S1.** Experimental setup of tip-enhanced SFG spectroscopy. Abbreviations are as follows, BS: beam splitter, OPO: optical parametric oscillator, FB etalon: Fabry-Pérot etalon, DM: dichroic mirror, CaF₂ AL: CaF₂ aspherical lens, SPF: short pass filter.

In experiments, the output from an amplified Yb-fiber laser (1033 nm, 280 fs, 40 W, 50 MHz; Monaco-1035-40-40, Coherent) was divided into two arms by a beam splitter. The first arm was used to pump a commercially available optical parametric oscillator (Levante IR, APE), and two pulses were generated: signal (1.3–2 μm) and idler (2.1–5 μm). We used the idler pulses to resonantly excite the CH-stretch vibrational modes of 4-methylbenzenethiol (4-MBT). The spectra of idler pulses are shown in Fig. S2(a). The second arm of the output from the Yb-fiber laser transmitted an air-spaced Fabry-Pérot etalon (SLS Optics) to make the spectral width narrow whose spectrum is shown in Fig. S2(b). The two laser pulses were combined collinearly at a dichroic mirror (custom made, Sigma Koki), and simultaneously incident on the nanojunction formed between the gold STM-tip and the sample substrate with the incident angle of 55°. The temporal overlap between two pulses was adjusted by a delay line in the second arm (1033 nm). These two pulses were linearly polarized along the tip axis (p-polarization) to effectively excite the gap-mode plasmon. The emitted sum frequency generation (SFG) light was detected with the reflection angle of 55°. We used two identical CaF₂ aspherical lenses (custom made, Natsume Optical Corporation) for focusing incident light and collecting SFG light. The resultant SFG signal was collimated by the CaF₂ aspherical lens, transmitted a short pass filter (FESH0950, Thorlabs), and then collected by an achromatic doublet (f = 50 mm; AC254-050-AB, Thorlabs) into a multi-mode optical fiber (Ø200 μm, NA = 0.22; M122L05, Thorlabs). Finally, the SFG signal transmitted after the optical fiber was dispersed in a spectrometer (Kymera 328i, Andor) and detected by an electronically cooled 2000 × 256 channels CCD detector (iDus416, Andor). The distance between the STM-tip and the substrate was controlled by a scanning tunneling microscope system (USM1400, Unisoku), by setting the tunnelling current to a constant value. Although our STM is available for use in an ultra-high-vacuum, low-temperature environment, we performed the experiment under atmospheric conditions at room temperature.

**Spectra of incident lasers.**

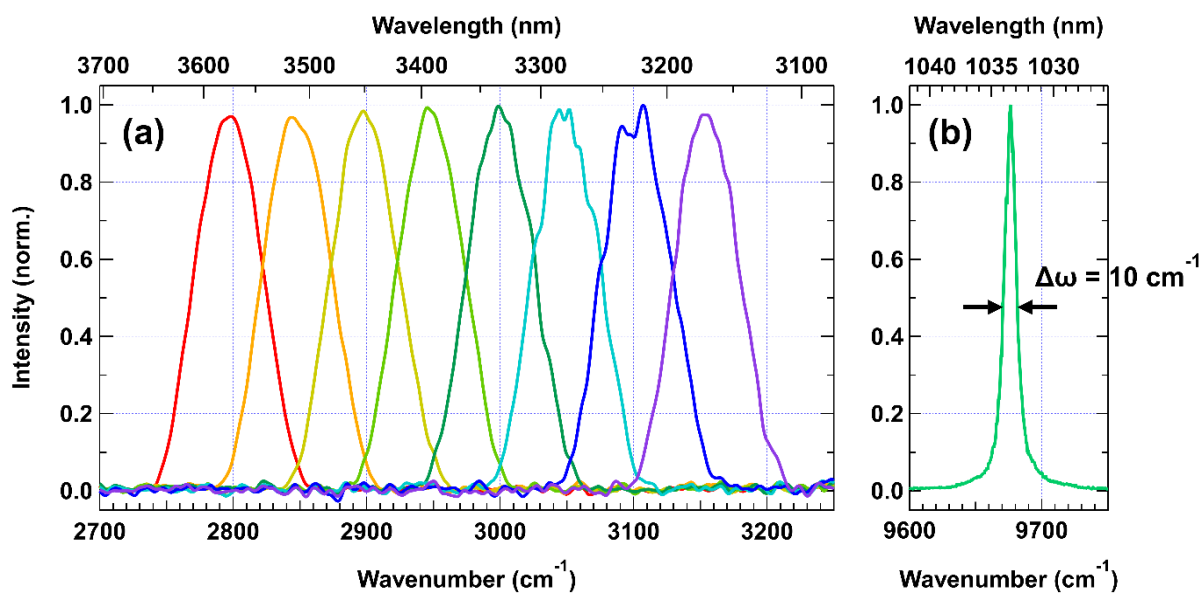

**Figure S2.** (a) The spectra of idler pulses with the central wavenumber changed from 2800 cm$^{-1}$ to 3150 cm$^{-1}$ by 50 cm$^{-1}$. These spectra were measured with FTIR spectrometer (FT-MIR, ARC Optics). (b) The spectrum of the output from the Yb-fiber laser transmitted after the Fabry-Pérot etalon. The spectrum was measured with the spectrometer described in Fig. S1. The full width at half maximum was 10 cm$^{-1}$, which determines the spectroscopic resolution of SFG spectra.

**Sample preparation of a self-assembled monolayer on a gold surface.** A 300-nm-thick gold thin film evaporated on a mica substrate was purchased from Unisoku and used for a base of a self-assembled monolayer (SAM) of 4-methylbenzenethiol (4-MBT) from Sigma-Aldrich. To prepare an atomically flat surface of Au(111), the substrate was flame-annealed, and then immersed in a 1 mM ethanol solution of 4-MBT for 24 hours. After that, the sample was rinsed with pure ethanol.

**SEM images of the Au STM tips.** SEM images of the Au STM tips used in the experiments are shown in Fig. S3: (a–d) for Au Tip #1 (used in the main text) and (e–h) for Au Tip #2 (used for reproducibility confirmation in the Supporting Information).

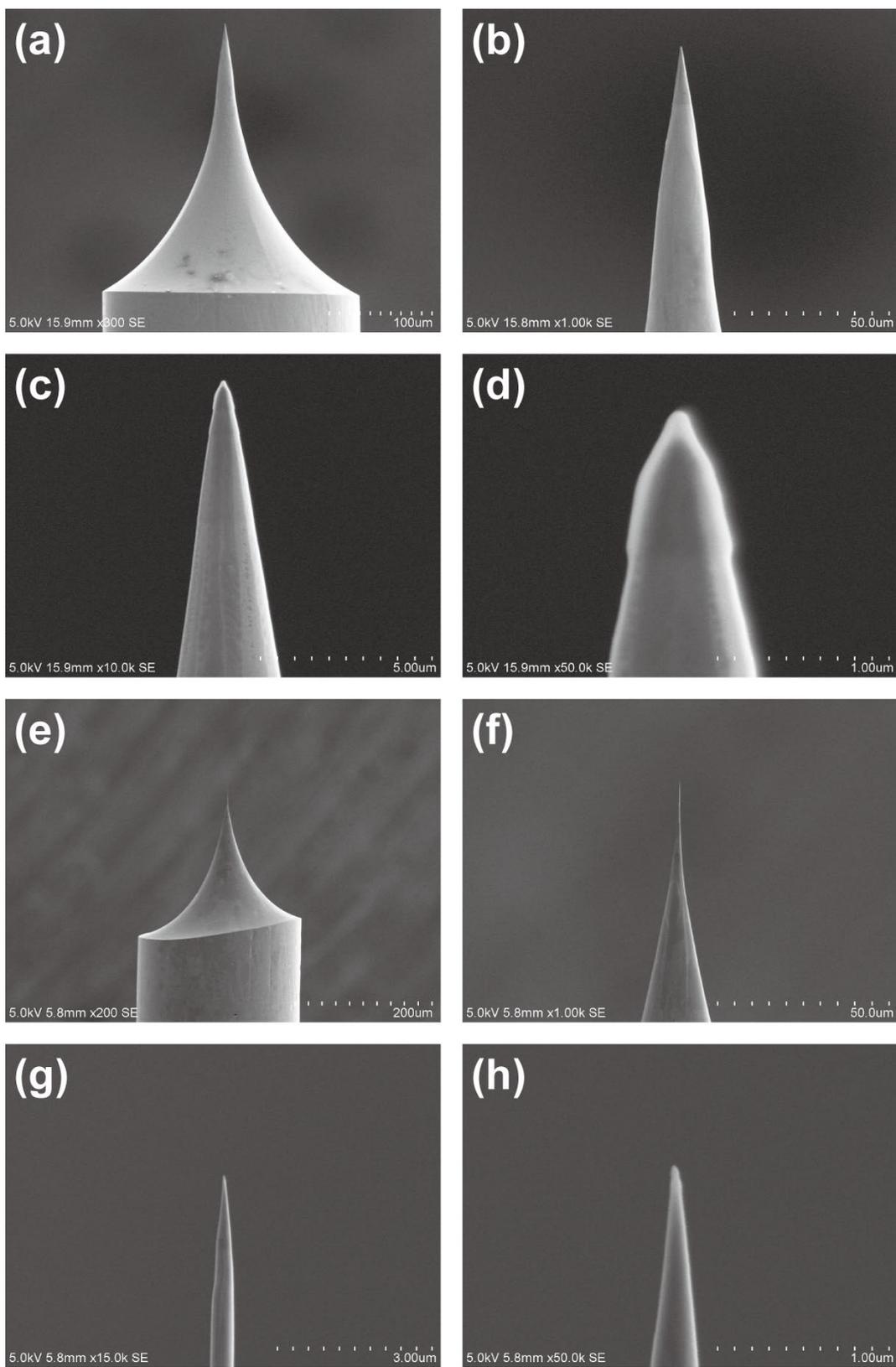

**Figure S3.** SEM images of the Au STM tips at various resolutions: (a–d) show Au Tip #1 used in the main text experiments, and (e–h) show Au Tip #2 used for reproducibility confirmation in the Supporting Information.

## 2. Interpretation of vibrational SFG spectra

The SFG intensity at frequency $\omega_{\text{SFG}} = \omega_{\text{vis}} + \omega_{\text{IR}}$ is given by[1]

$$I(\omega_{\text{SFG}}) \propto \left| \chi^{(2)}(\omega_{\text{SFG}}; \omega_{\text{vis}}, \omega_{\text{IR}}) \right|^2 I_0(\omega_{\text{vis}}) I_0(\omega_{\text{IR}}), \qquad (\text{S.1})$$

where $\chi^{(2)}(\omega_{\text{SFG}}; \omega_{\text{vis}}, \omega_{\text{IR}})$ is the second-order nonlinear susceptibility, and $I_0(\omega_{\text{vis}})$ and $I_0(\omega_{\text{IR}})$ are the intensities of the incident visible (or near-IR) and IR pulses, respectively. When $\omega_{\text{IR}}$ is in the mid-IR region, associated with vibrational resonances, and $\omega_{\text{vis}}$ is in the visible or near-IR region, far below the electronic resonance, $\chi^{(2)}(\omega_{\text{SFG}})$ is expressed as follows:[2-8]

$$\chi^{(2)}(\omega_{\text{IR}}) = \chi_{\text{NR}}^{(2)} + \chi_{\text{R}}^{(2)}(\omega_{\text{IR}}) = A_{\text{NR}} e^{i\theta} + \sum_q \frac{A_q}{\omega_q - \omega_{\text{IR}} - i\Gamma_q}. \qquad (\text{S.2})$$

In this equation, $\chi_{\text{NR}}^{(2)}$ and $\chi_{\text{R}}^{(2)}(\omega_{\text{IR}})$ represent the vibrational non-resonant and vibrational resonant contributions, respectively, and the frequency is converted from $\omega_{\text{SFG}}$ to $\omega_{\text{IR}}$. Here, $A_q$, $\omega_q$, and $\Gamma_q$ are the amplitude, resonant frequency, and damping constant of the $q$-th vibrational mode, respectively. The amplitude $A_q$ is given by[4, 5, 8-10]

$$A_q = \frac{1}{2\epsilon_0 \omega_q} \left( \frac{\partial \alpha^{(1)}}{\partial Q_q} \right) \left( \frac{\partial \mu}{\partial Q_q} \right), \qquad (\text{S.3})$$

where $\epsilon_0$ represents the permittivity of vacuum, $\alpha^{(1)}$ and $\mu$ represent the Raman polarizability and the IR transition dipole moment, respectively, and $Q_q$ is the normal coordinate of the $q$-th vibrational mode. Eq. (S.3) indicates that vibrational SFG is a nonlinear optical process induced by IR absorption and subsequent anti-Stokes Raman scattering. The non-resonant susceptibility $\chi_{\text{NR}}^{(2)}$ is characterized by a constant amplitude $A_{\text{NR}}$ with a phase factor $\theta$. In general, the term $\chi_{\text{NR}}^{(2)}$ should be small and real when the substrate is a dielectric material and does not have resonances with light. However, for metal or semiconductor substrates, $\chi_{\text{NR}}^{(2)}$ is generally complex and not small at all.[5, 6]

If the IR pulse has a broad spectrum, as illustrated in Fig. S2, it can be approximated by a Gaussian function with central frequency $\omega_0$ and spectral width $\sigma$:

$$I_0(\omega_{\text{IR}}) \propto \exp \left[ -\frac{(\omega_{\text{IR}} - \omega_0)^2}{\sigma^2} \right], \qquad (\text{S.4})$$

and if the visible pulse has a narrowband spectrum, also as illustrated in Fig. S2, then Eq. (S.1) becomes:[7]

$$I(\omega_{IR}) = C \exp\left[-\frac{(\omega_{IR} - \omega_0)^2}{\sigma^2}\right]\left|A_{NR}e^{i\theta} + \sum_q \frac{A_q}{\omega_q - \omega_{IR} - i\Gamma_q}\right|^2. \qquad (S.5)$$

Here, $C$ represents the relative intensity of the SFG signal, and the frequency is converted from $\omega_{SFG}$ to $\omega_{IR}$, as in Eq. (S2). To reduce the number of parameters, we additionally modify Eq. (S.5) as follows:

$$I(\omega_{IR}) = c \exp\left[-\frac{(\omega_{IR} - \omega_0)^2}{\sigma^2}\right]\left|1 + \sum_q \frac{r_q e^{-i\theta}}{\omega_q - \omega_{IR} - i\Gamma_q}\right|^2, \qquad (S.6)$$

where $r_q = A_q/A_{NR}$ and $c = CA_{NR}^2$. This equation is given as Eq. (1) in the main text.

## 3. Far-field SFG measurements and analysis

We performed SFG measurements in a far-field geometry using the optical setup described in the prvious section (1. Methods). During the measurement, the tip–sample distance was maintained at more than 1 μm away to avoid the gap-mode plasmon resonance. The input powers of the IR and 1033 nm lasers were set to 50 mW and 5 mW, respectively, and the measurement time was 5 minutes. We observed three dip structures in the SFG spectra, as shown in Fig. S4. Two of these spectra were smoothed and fitted using Eq. (S.6), assuming two vibrational modes within each spectral range. The fitting results are shown as the black dashed curves in Fig. S5. The red curves represent the imaginary part of the vibrational resonant contribution in the second-order susceptibility $(\mathrm{Im}\left[\chi_R^{(2)}(\omega_{IR})\right])$, which corresponds to the purely absorptive vibrational responses given by the following equation.

$$\mathrm{Im}\left[\chi_R^{(2)}(\omega_{IR})\right] \propto \mathrm{Im}\left[\sum_{q=1}^2 \frac{r_q}{\omega_q - \omega_{IR} - i\Gamma_q}\right] = \sum_{q=1}^2 \frac{r_q\Gamma_q}{\left(\omega_q - \omega_{IR}\right)^2 + \Gamma_q^2}.$$

In Fig. S5, the best-fitting frequencies of the vibrational modes are 2856 and 2903 cm⁻¹ in (a), and 2902 and 2952 cm⁻¹ in (b). These frequencies correspond to a methyl symmetric stretching (CH₃-SS) vibration, a Fermi resonance (CH₃-FR) between the CH₃-SS and an overtone of a methyl bending vibration, and a degenerate methyl stretching (CH₃-DS) vibration.[11] The best-fitting parameters used in Fig. S5 are listed in Table S1.

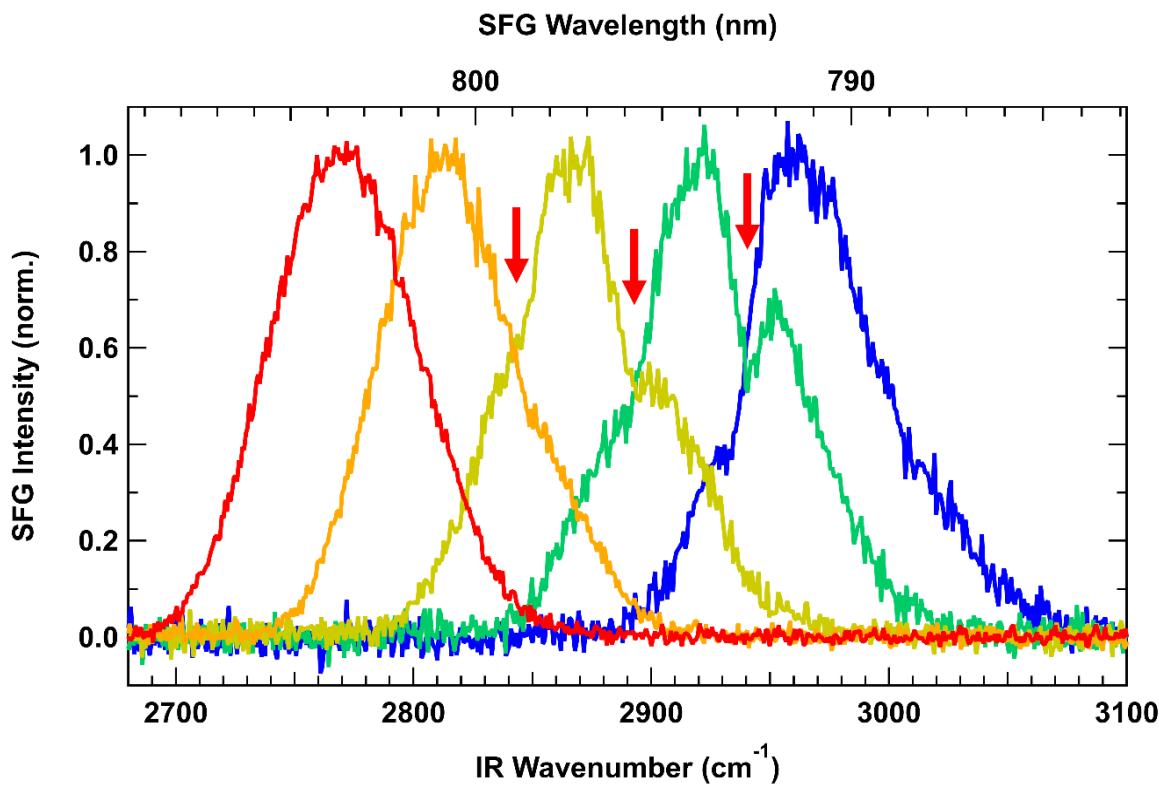

**Figure S4.** The observed SFG spectra in a far-field geometry. Dip structures were observed at the positions indicated by the red arrows.

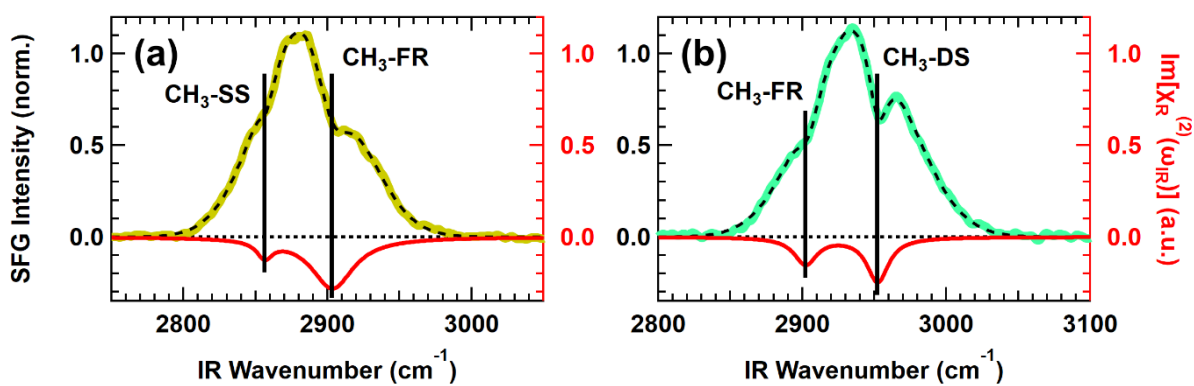

**Figure S5.** SFG spectra in Fig. S4 were smoothed and fitted using Eq. (S.6), assuming two vibrational modes within each spectral range. The fitting results are shown by the black dashed curves. The red curves represent $\mathrm{Im}\left[\chi_{\mathrm{R}}^{(2)}(\omega_{\mathrm{IR}})\right]$, which corresponds to the purely absorptive vibrational responses. The extracted peaks are (a) CH₃-SS at 2856 cm⁻¹ and CH₃-

FR at 2903 cm$^{-1}$ and (b) CH$_3$-FR at 2902 cm$^{-1}$ and CH$_3$-DS at 2952 cm$^{-1}$. The left axis corresponds to the normalized intensity of SFG, and the right axis corresponds to the intensity of the extracted $\mathrm{Im}\left[\chi_{\mathrm{R}}^{(2)}(\omega_{\mathrm{IR}})\right]$, with both axes having the same scale.

**Table S1.** Fitting parameters used in Fig. S5.

|  | (a) | (b) |
|---|---|---|
| $c$ | $1.35 \pm 0.02$ | $1.25 \pm 0.01$ |
| $\omega_0$ (cm$^{-1}$) | $2887.8 \pm 0.2$ | $2938.8 \pm 0.2$ |
| $\sigma$ (cm$^{-1}$) | $44.6 \pm 0.2$ | $48.0 \pm 0.1$ |
| $\theta$ (deg) | $89.74 \pm 0.04$ | $89.71 \pm 0.03$ |
| $r_{q_1}$ | $-0.61 \pm 0.06$ | $-1.41 \pm 0.10$ |
| $\omega_{q_1}$ (cm$^{-1}$) | $2856.4 \pm 0.3$ | $2902.2 \pm 0.3$ |
| $\Gamma_{q_1}$ (cm$^{-1}$) | $6.61 \pm 0.58$ | $9.61 \pm 0.56$ |
| $r_{q_2}$ | $-5.01 \pm 0.21$ | $-2.25 \pm 0.06$ |
| $\omega_{q_2}$ (cm$^{-1}$) | $2903.4 \pm 0.5$ | $2952.3 \pm 0.2$ |
| $\Gamma_{q_2}$ (cm$^{-1}$) | $17.88 \pm 0.53$ | $9.21 \pm 0.23$ |

## 4. Enhancement mechanism in TE-SFG spectroscopy

The tip-enhanced SFG (TE-SFG) process consists of three steps: (i) electric field enhancement in the nanogap for incident IR and visible light; (ii) induction of second-order polarization by an enhanced IR and visible electric field; and (iii) radiation of sum frequency light. In the first step, the electric field of incident light, $E_0(\omega)$, is plasmonically enhanced in the nanogap. We define the enhancement factor $K_{\mathrm{gap}}(\omega)$ as the ratio of the incident field to the enhanced electric near field, $E_{\mathrm{gap}}(\omega)$, as[6, 12, 13]

$$E_{\mathrm{gap}}(\omega) = K_{\mathrm{gap}}(\omega)E_0(\omega). \tag{S.7}$$

This enhanced field induces the second-order polarization $P^{(2)}(\omega_{\mathrm{SFG}} = \omega_{\mathrm{vis}} + \omega_{\mathrm{IR}})$ as

$$
\begin{aligned}
P^{(2)}(\omega_{\mathrm{SFG}}) &= \epsilon_0 \chi^{(2)}(\omega_{\mathrm{SFG}}; \omega_{\mathrm{vis}}, \omega_{\mathrm{IR}})E_{\mathrm{gap}}(\omega_{\mathrm{vis}})E_{\mathrm{gap}}(\omega_{\mathrm{IR}}) \\
&= \epsilon_0 \chi^{(2)}(\omega_{\mathrm{SFG}}; \omega_{\mathrm{vis}}, \omega_{\mathrm{IR}})\left[K_{\mathrm{gap}}(\omega_{\mathrm{vis}})E_0(\omega_{\mathrm{vis}})\right]\left[K_{\mathrm{gap}}(\omega_{\mathrm{IR}})E_0(\omega_{\mathrm{IR}})\right].
\end{aligned} \tag{S.8}
$$

The induced polarization subsequently emits a sum frequency field as

$$E_{\mathrm{SFG}}(\omega_{\mathrm{SFG}}) \propto L_{\mathrm{gap}}(\omega_{\mathrm{SFG}})P^{(2)}(\omega_{\mathrm{SFG}}), \tag{S.9}$$

where $L_{\text{gap}}(\omega_{\text{SFG}})$ is the radiation efficiency of TE-SFG from $P^{(2)}(\omega_{\text{SFG}})$ in the nanogap.[6, 12] The resultant TE-SFG intensity $I(\omega_{\text{SFG}})$ is given by

$$I(\omega_{\text{SFG}}) \propto |E_{\text{SFG}}(\omega_{\text{SFG}})|^2$$

$$\propto |L_{\text{gap}}(\omega_{\text{SFG}})|^2 |K_{\text{gap}}(\omega_{\text{vis}})|^2 |K_{\text{gap}}(\omega_{\text{IR}})|^2 |\chi^{(2)}(\omega_{\text{SFG}}; \omega_{\text{vis}}, \omega_{\text{IR}})|^2 I_0(\omega_{\text{vis}}) I_0(\omega_{\text{IR}}), \quad \text{(S.10)}$$

where $I_0(\omega)$ is the intensity of the incident light $(I_0(\omega) \propto |E_0(\omega)|^2)$. The comparison between Eqs. (S.1) and (S.10) shows that the SFG enhancement is governed by the factor of $|L_{\text{gap}}(\omega_{\text{SFG}})|^2 |K_{\text{gap}}(\omega_{\text{vis}})|^2 |K_{\text{gap}}(\omega_{\text{IR}})|^2$, while the second-order susceptibility $\chi^{(2)}(\omega_{\text{SFG}}; \omega_{\text{vis}}, \omega_{\text{IR}})$ is independent of electric field enhancement. Therefore, TE-SFG spectra can be interpreted similarly to far-field observations, except for the influence of electric field enhancement.

# 5. Numerical procedure for electromagnetic field simulations

We performed electromagnetic (EM) field simulations using the finite-difference time-domain method with commercial software (Lumerical FDTD, Ansys). In these simulations, we modeled the STM tip as a rounded cone of gold, positioned perpendicular to the gold substrate with a tip-to-substrate distance of 1 nm. The refractive index of gold was taken from the experimental values of Olmon et al.[14] To evaluate the incident field enhancement, denoted as $K_{\text{gap}}(\omega)$ in §4, we placed a monitor at the midpoint between the tip apex and the substrate surface to measure the EM field strength. The incident light was modeled as a Gaussian beam focused on the nanojunction between the tip and the substrate at an incident angle of 55°, with a waist radius of 2 μm, to simulate the experimental geometry. The resulting field enhancement values are displayed as a function of incident light wavelength in Fig. 1(a) of the main text.

To evaluate the radiation efficiency, denoted as $L_{\text{gap}}(\omega)$ in §4, a dipole moment perpendicular to the gold substrate was placed at the same position as the monitor, and the radiated EM field from the dipole was monitored as the far-field at a position where the lateral and vertical distances from the dipole were 3000 and 2100 nm, respectively, matching the reflection angle of 55°. The obtained radiation efficiency values are displayed as a function of incident light wavelength in Fig. 1(b) of the main text.

## 6. Best-fitting parameters used in Figure 4

**Table S2.** Fitting parameters used in Figure 4 in the main text

|  | (a) | (b) | (c) |
|---|---|---|---|
| $c$ | $1.29 \pm 0.02$ | $1.76 \pm 0.03$ | $1.65 \pm 0.03$ |
| $\omega_0$ (cm$^{-1}$) | $2848.7 \pm 0.1$ | $2896.3 \pm 0.2$ | $2948.7 \pm 0.2$ |
| $\sigma$ (cm$^{-1}$) | $37.5 \pm 0.2$ | $35.8 \pm 0.2$ | $34.3 \pm 0.2$ |
| $\theta$ (deg) | $99.6 \pm 2.7$ | $70.3 \pm 1.9$ | $108.4 \pm 2.1$ |
| $r_q$ | $-2.62 \pm 0.14$ | $-6.05 \pm 0.27$ | $-5.25 \pm 0.25$ |
| $\omega_q$ (cm$^{-1}$) | $2852.5 \pm 0.4$ | $2904.2 \pm 0.4$ | $2939.2 \pm 0.5$ |
| $\Gamma_q$ (cm$^{-1}$) | $12.4 \pm 0.5$ | $16.1 \pm 0.5$ | $16.3 \pm 0.5$ |

## 7. Confirmation of the reproducibility of dip structures in TE-SFG spectra

We conducted experiments with varying parameters to confirm the reproducibility of dip structures caused by vibrational resonances in TE-SFG spectra. In Fig. S6, the spectra in (a) and (b) were obtained using Au Tip #1 (shown in Fig. S3), with the bias voltage and tunneling current set to 0.5 V and 0.5 nA, respectively, and with the input powers of the IR and 1033 nm lasers set at (a) 20 mW and 0.08 mW, and (b) 50 mW and 0.5 mW. The spectra in (c) were obtained using Au Tip #2 (also shown in Fig. S3), with the bias voltage and tunneling current set to 0.05 V and 2 nA, respectively, and with the input powers of the IR and 1033 nm lasers set at 20 mW and 0.5 mW, respectively. The measurement time for each spectrum was 10 minutes across all experiments.

In the spectra in Figs. S6(a–c), three dip structures are observed in the TE-SFG spectra at the positions indicated by the red arrows. Despite small differences in appearance, the reproducibility of dip structures associated with vibrational resonances in TE-SFG spectra has been confirmed. While the input powers in (b) were higher than in (a), the signal intensities in (a) exceeded those in (b). The spectra in (b) were measured one day after those in (a), during which the tip apex was likely altered, resulting in weakened gap-mode plasmon resonance. Additionally, due to the relatively strong input powers in the measurement in Fig. S6(b), weak far-field SFG signals were observed even though the sample substrate was retracted 50 nm from the tip apex, as represented by the black curves in Fig. S7. However, the intensities of these far-field SFG signals were considerably weaker than those of the TE-SFG signals. Therefore, we conclude that the spectra in Fig. S6 are primarily due to the near-field contribution.

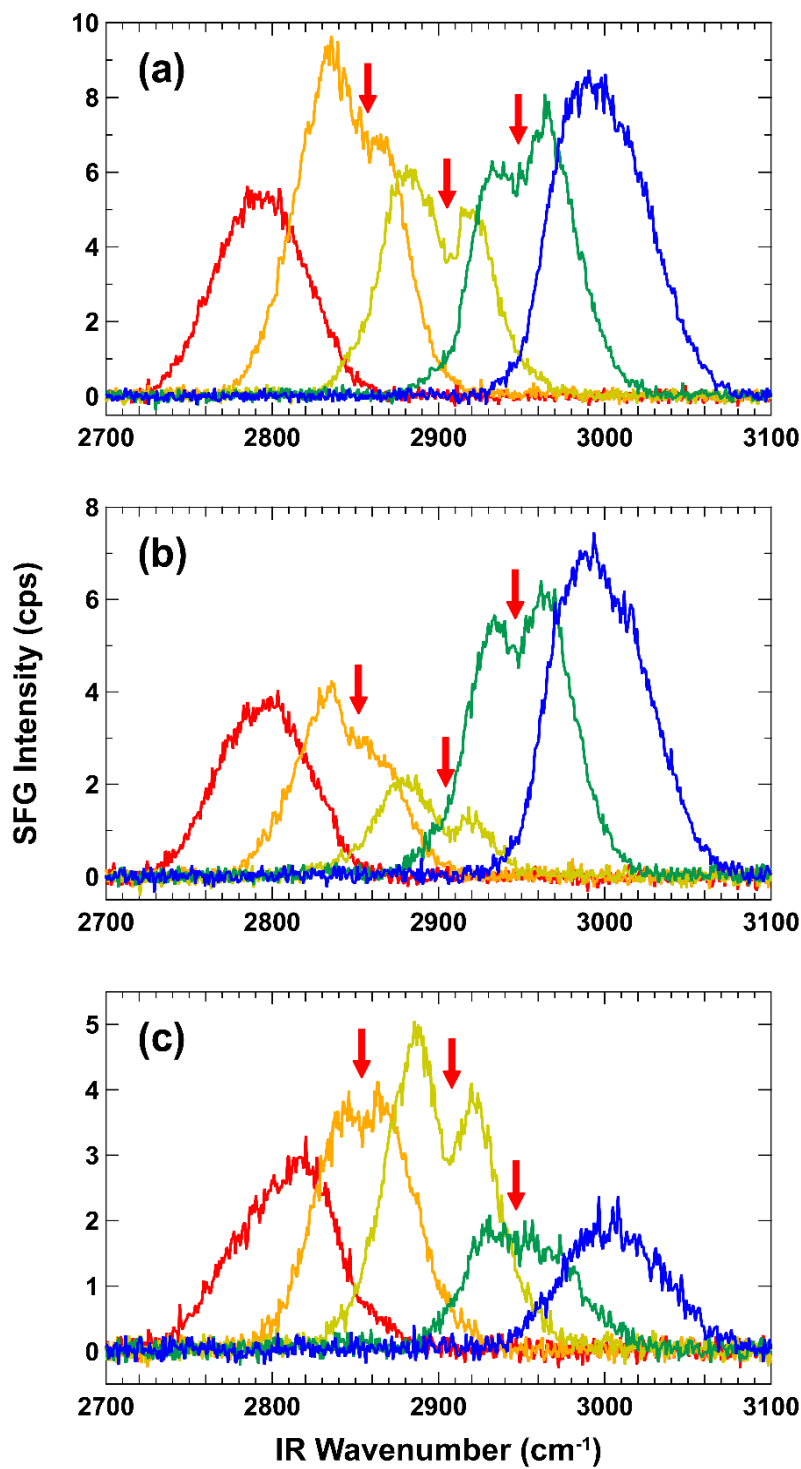

**Figure S6.** TE-SFG spectra measured under different experimental conditions. The parameters were (a) IR power: 20 mW, 1033 nm power: 0.08 mW, bias voltage: 0.5 V, tunneling current: 0.5 nA, with Au Tip #1; (b) IR power: 50 mW, 1033 nm power: 0.5 mW, bias voltage: 0.5 V, tunneling current: 0.5 nA, with Au Tip #1; (c) IR power: 20 mW, 1033 nm power: 0.5 mW, bias voltage: 0.05 V, tunneling current: 2 nA, with Au Tip #2. The measurement time for each spectrum was 10 minutes across all experiments. Dip structures were observed at the position indicated by the red arrows.

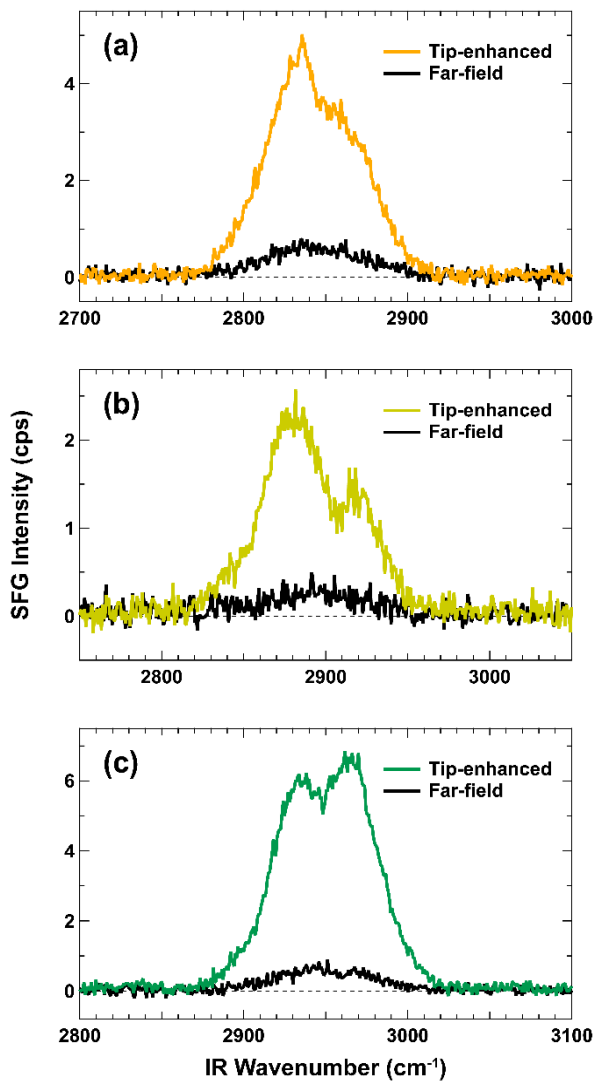

**Figure S7.** Comparison of tip-enhanced and far-field SFG spectra measured under the experimental conditions described in Fig. S6(b). The colored curves were obtained with a tunneling current of 0.5 nA and a bias voltage of 0.5 V (Tip-enhanced, Au Tip #1), while the black curves were obtained with the sample substrate retracted by 50 nm (Far-field). The spectra in Fig. S6(b) are the tip-enhanced spectra with the far-field spectra subtracted as background.

## 8. Measurement procedure for the approach curve of TE-SFG

In the approach curve measurement of TE-SFG, we observed the TE-SFG spectra by varying the relative tip–sample distance, which was controlled by adjusting the tunneling current. The bias voltage was set to 0.5 V, and we defined the condition in which the tunneling current was 5 nA as the zero position of the relative tip–sample distance, since the absolute value of the tip–sample distance could not be strictly determined experimentally. The dependence of the tunneling current on the relative tip–sample distance (I-Z curve) was measured for a self-assembled monolayer (SAM) of 4-methylbenzenethiol (4-MBT) on an Au(111) film using the Z-spectroscopy function in an STM controller (Nanonis, SPECS), without laser irradiation. The tunneling current was recorded sequentially as the relative tip–sample distance was increased. This measurement was repeated 10 times to calculate the average, which was shown in Fig. S8. The correlation between the relative tip–sample distance and the tunneling current is displayed in Table S3, from which we set the tunneling current for specified relative tip–sample distances. We note that while the I-Z curve fits an exponential function for Z > 0.6 nm, it deviates from this behavior for Z < 0.6 nm. This deviation is likely due to contact between the tip apex and the top surface of the SAM, as observed in previous research.[15] In the TE-SFG spectral measurements, the input powers of the IR and 1033 nm lasers were 20 and 0.2 mW, respectively, with an exposure time of 5 s and 12 repetitions (totaling 1 minute). To avoid sample damage, the sample was scanned over a 100 nm × 100 nm area with a scanning speed of 700 nm/s.

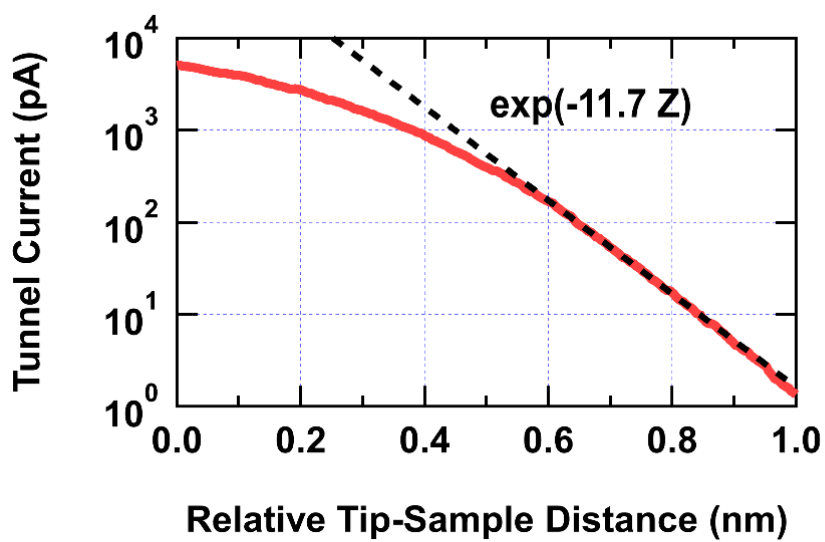

**Figure S8.** The dependence of the tunneling current on the relative tip–sample distance (I-Z curve) was measured for a SAM of 4-MBT on a Au(111) film at a bias of 0.5 V without laser irradiation.

**Table S3.** Correlation between the relative tip–sample distance and the tunneling current

| Relative tip–sample distance (nm) | Tunneling current (pA) |
|---|---|
| 0.00 | 5066 |
| 0.16 | 3049 |
| 0.33 | 1390 |
| 0.49 | 429 |
| 0.65 | 92.1 |
| 0.82 | 14.0 |
| 0.98 | 1.72 |

## 9. Temporal variation in tip–sample distance and tunneling current before and during laser irradiation

We recorded the temporal variations in the Z position of the sample and the tunneling current before and during laser irradiation, with the tunneling current set to 1.72 pA (Figs. S9(a, c)) and 5.066 nA (Figs. S9(b, d)). These current values correspond to the relative tip–sample distances of 0.98 nm and 0.00 nm, respectively (Table S3). The input powers of the IR and 1033-nm lasers were 20 mW and 0.2 mW, respectively, consistent with the values used in measuring the TE-SFG approach curve (Fig. 5(b) of the main text). The time origin was defined as the moment when laser irradiation began. The positive Z direction indicates the sample substrate moving away from the tip. In both tunneling current cases, the Z position shifted rapidly by ~2 nm immediately after laser irradiation, indicating an increase in the tip–sample distance (Figs. S9(a,b)). However, these rapid shifts subsided within 5 seconds. In the actual measurement of the TE-SFG approach curve, we began laser irradiation with the sample substrate positioned away from the tip and waited for over a minute before moving the sample closer to the tip. Thus, the rapid shifting effect was likely mitigated. During laser irradiation, spikes in the tunneling current occurred more frequently (Figs. S9(c, d)), and the fluctuations in the Z position became pronounced (Figs. S9(a, b)). Accordingly, the standard deviations of the Z position increased from 0.17 nm to 0.42 nm at 1.72 pA and from 0.16 nm to 0.54 nm at 5.066 nA (Table S4). These standard deviations of the Z position during laser irradiation were used as the error bars for the relative tip–sample distance in the TE-SFG approach curve (Fig. 5(b) of the main text). The current spikes likely result from plasmon-assisted tunneling, where electron tunneling is driven by plasmonic field enhancement.[16] The current spikes are more pronounced at 5.066 nA due to the shorter tip–sample distance and stronger plasmonic enhancement (Fig. S9(d)). However, since the fluctuations in the tip–sample distance remained within a few nanometers, we conclude that the nanogap between the tip and the substrate, which serves as the signal-generation region, is also at most a few nanometers.

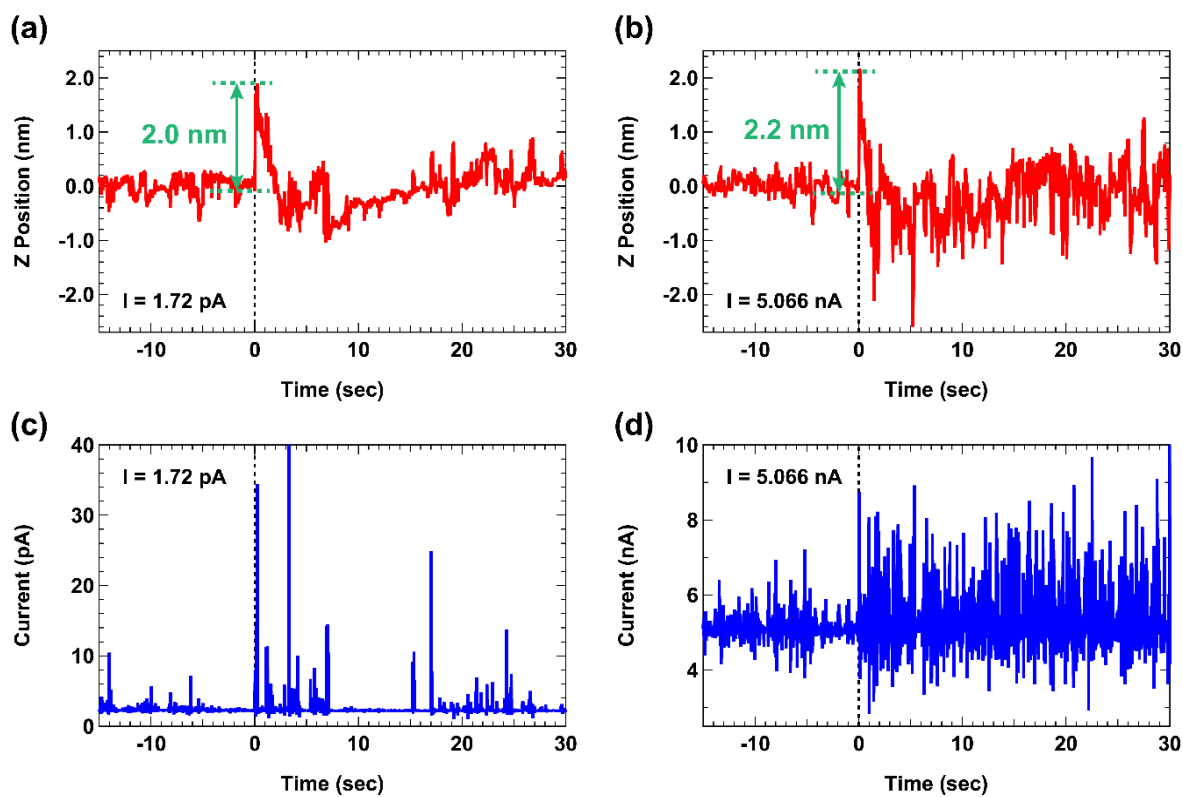

**Figure S9.** Red curves: temporal variations in the Z position of the sample before and during laser irradiation, with the tunneling current set to (a) 1.72 pA and (b) 5.066 nA. Immediately after laser irradiation, the Z position shifted rapidly by 2.0 nm at 1.72 pA and 2.2 nm at 5.066 nA, but these shifts subsided within 5 seconds. Blue curves: temporal variations in the tunneling current before and during laser irradiation, with the tunneling current set to (c) 1.72 pA and (d) 5.066 nA.

**Table S4.** The average values and standard deviations of the Z position and tunneling current before and during laser irradiation, with the tunneling current set to 1.72 pA and 5.066 nA.

| | Before laser irradiation (-15–0 seconds) | | During laser irradiation (0–30 seconds) | |
|---|---|---|---|---|
| Set current | Z position | Current | Z position | Current |
| 1.72 pA | 0 ± 0.17 nm | 2.36 ± 0.50 pA | -0.02 ± 0.42 nm | 2.50 ± 1.95 pA |
| 5.066 nA | 0 ± 0.16 nm | 5.12 ± 0.31 nA | -0.16 ± 0.54 nm | 5.33 ± 0.81 nA |

## 10. Intensity fluctuations in TE-SFG signals for each relative tip–sample distance

In the approach curve measurement of TE-SFG, we repeated the measurement 12 times with an exposure time of 5 seconds for each relative tip–sample distance. The SFG signal intensities fluctuated with each measurement number, and correspondingly, the integrated SFG counts for each relative tip–sample distance also fluctuated, as shown in Fig. S10. As the relative tip–sample distance decreased, both the integrated SFG counts and their fluctuations increased. Table S5 lists the average values and standard deviations of the integrated SFG counts for each relative tip–sample distance. These values are used in Fig. 4(b) in the main text.

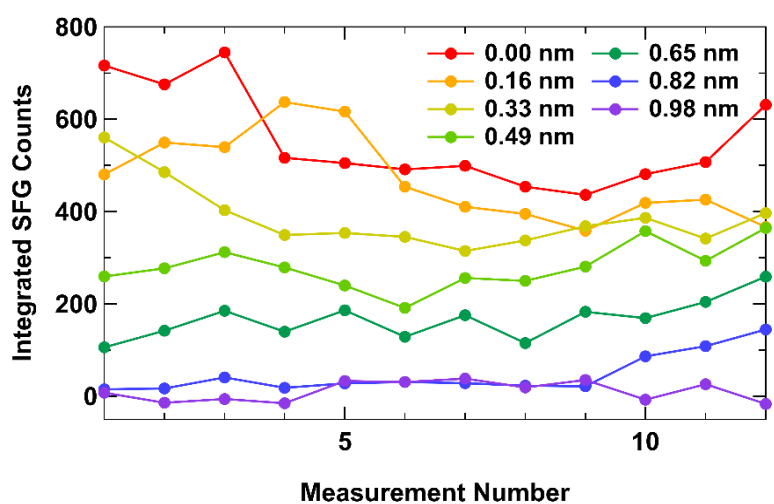

**Figure S10.** Integrated SFG counts for each relative tip–sample distance as a function of measurement number. The relative tip–sample distance varied from 0.00 to 0.98 nm.

**Table S5.** The average values and standard deviations of the integrated SFG counts for each relative tip–sample distance.

| Relative tip–sample distance (nm) | Integrated SFG counts |
|---|---|
| 0.00 | $555 \pm 107$ |
| 0.16 | $471 \pm 94$ |
| 0.33 | $387 \pm 70$ |
| 0.49 | $280 \pm 48$ |
| 0.65 | $166 \pm 43$ |
| 0.82 | $47 \pm 42$ |
| 0.98 | $11 \pm 21$ |